\documentclass[a4paper]{article}
\usepackage{amsfonts}
\usepackage{amsmath}
\usepackage{amssymb}
\usepackage{geometry}
\usepackage{graphicx}
\usepackage{authblk}
\usepackage{caption}
\usepackage{hyperref}
\hypersetup{colorlinks = false, urlbordercolor = {0 1 0}}
\usepackage{cite}
\title{Whirligig Beetles as Corralled Active Brownian Particles}
\author[1,2]{Harvey L. Devereux}
\author[3]{Colin R. Twomey}
\author[4,5]{Matthew S. Turner}
\author[2,6]{Shashi Thutupalli}
\affil[1]{Department of Mathematics, University of Warwick, Coventry CV4 7AL, UK}
\affil[2]{Simons Center for the Study of Living Machines, National Centre for Biological Sciences,Tata Institute for Fundamental Research, Bangalore 560065, India.}
\affil[3]{Department of Biology, and Mind Center for Outreach, Research, and Education, University of Pennsylvania, Philadelphia, PA, USA}
\affil[4]{Department of Physics and}\affil[4]{Centre for Complexity Science, University of Warwick, Coventry CV4 7AL, UK}
\affil[5]{Department of Chemical Engineering, Kyoto University, Kyoto, 615-8510, Japan}
\affil[6]{International Centre for Theoretical Sciences, Tata Institute for Fundamental Research, Bangalore 560089, India}
\date{\today}
\providecommand{\keywords}[1]
{
  \small
  \textbf{\textit{Keywords---}} #1
}

\begin{document}

\maketitle
\begin{abstract}
We study the collective dynamics of groups of whirligig beetles \textit{Dineutus discolor (Coleoptera: Gyrinidae)} swimming freely on the surface of water. We extract individual trajectories for each beetle, including positions and orientations, and use this to discover (i) a density dependent speed scaling like $v\sim\rho^{-\nu}$ with $\nu\approx0.4$ over two orders of magnitude in density
 (ii) an {inertial delay} for velocity alignment of $\sim13$ ms and (iii) coexisting high and low density phases, consistent with motility induced phase separation (MIPS). We modify a standard 
 active brownian particle (ABP) model to a Corralled ABP (CABP) model that functions in open space by incorporating a density-dependent reorientation of the beetles, towards the cluster. We use our new model to test our hypothesis that a MIPS (or a MIPS like effect) can explain the co-occurrence of high and low density phases we see in our data. The fitted model then successfully recovers a MIPS-like condensed phase for $N=200$ and the absence of such a phase for smaller group sizes $N=50,100$.
\end{abstract}
\\\\
\maketitle
\section{Introduction}
\begin{figure}[ht!]
\centering
\includegraphics[width=0.99\linewidth]{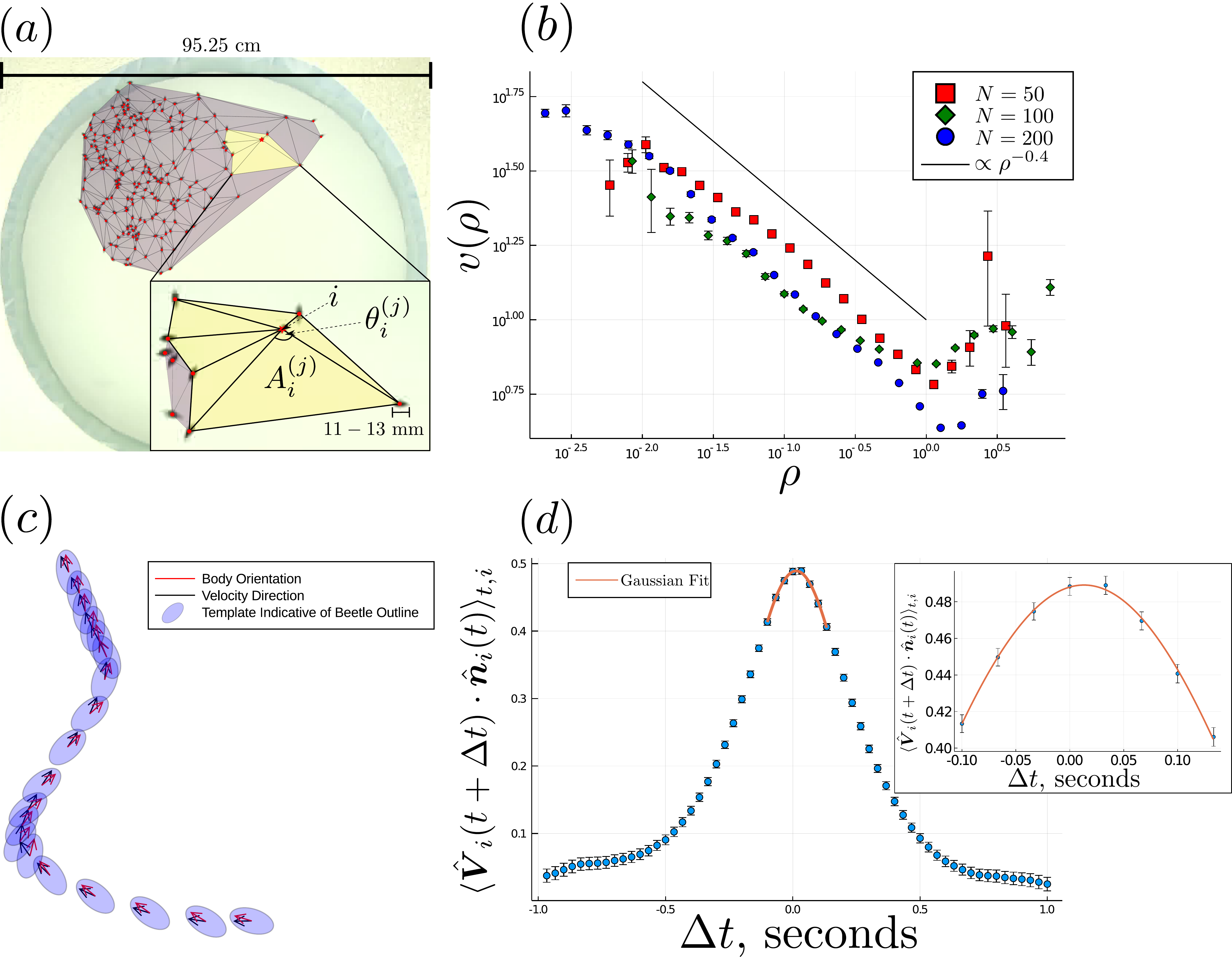}
\caption{ Beetle velocities scale with a power law of local density and exhibit a relatively short inertial delay of 13ms. $(a)$ A snapshot of the experimental setup and an overlay detailing the method for local density calculation using the Delaunay Tessellation based method (see methods section \ref{sec:dense}). In the overlay, the red points are particle (beetle) positions and lines indicate the Delaunay tessellation. The inset labels the angles and areas used to calculate the local density of the central $i^{\rm th}$ beetle, shown as a red star. The yellow polygon outlined in bold  indicates the union of Delaunay triangles having this beetle as a common vertex. We label this set of Delaunay triangles with the index
$j$ referring to each triangle as $T_{i}^{(j)}$, it's area is $A_{i}^{(j)}$, and the angle subtended at $i$ as $\theta_{i}^{(j)}$. The local density is calculated as the \textit{inverse} of the average weighted areas $A_{i}^{(j)}$, with the angles $\theta_{i}^{(j)}$ as weights, further normalised
by a factor of $1/2\pi$ to account for the fact internal points satisfy $\sum_{j}\theta_{i}^{(j)}=2\pi$ while points on the boundary satisfy $\sum_{j}\theta_{i}^{(j)}<2\pi$.
See methods section for further details.
$(b)$ The relationship between beetle speed $v$ (body lengths per second) and local density $\rho$ (per body length squared) on a log-log scale. Each data point represents a bin-average with error bars showing one standard deviation. The solid black line indicates a power law ($\rho^{-0.4}$) as a guide to the eye.
$(c)$ Shows the inertial delay between a beetle's velocity vector and its body axis orientation: the orientation leads the velocity. The shaded ellipses represent the moving outline of the beetle over time. This is quantified in
$(d)$, showing the orientation-velocity time correlation function with a Gaussian fit superimposed near the peak, located at $\Delta t = 13$ ms (see inset).
Only in panel $(d)$ we use the set of all $N=200$ beetle trajectories pre-filtered to remove (near) collisions (see SI for details).}
\label{fig:Panel1}
\end{figure}

There is now an extensive body of computer simulations and theoretical work to suggest that aggregation can emerge, even when inter-particle interactions are purely repulsive, e.g. steric contact forces \cite{Fily2012Athermal,Palacci2013living,Buttinonu2013Dynamical,redner2013a,Cates2008statistical}.
The aggregation can be thought of as arising from the competition between the accretion of free motile particles on contact with a dense cluster and their departure from the surface of that cluster following a re-orientational time lag. A form of non-equilibrium phase separation can arise in which densely and sparsely populated regions co-exist. This is now known as motility induced phase separation (MIPS). This phase separation depends on the relationship between self-propulsion speed and local density. If the speed falls off sharply enough with increasing density then a feedback loop can emerge in which slowing down (due to higher density) promotes further aggregation. High density clusters then grow and the density in the dilute phase drops. As it does so the rate of accretion onto the clusters drops until it again comes into balance with the evaporation of particles from the cluster surface into the dilute phase. Even steric repulsion is therefore enough to cause active, self propelled particles to accumulate in regions where they move slowly \cite{Schnitzer1993Theory}. MIPS has been shown to arise in active particle systems with a more general density dependent propulsion speed\cite{Cates2015Motility}. Our study provides experimental evidence justifying the use of a power law velocity-density dependence in such models.
\\\\
This aggregation of motile particles in the presence of purely repulsive interactions has attracted much attention from theorists working on non-equilibrium statistical mechanics attracted by possible  insights to time-reversal symmetries and entropy production. Numerical investigations of MIPS have been carried out to examine the affects of incorporating velocity alignment terms, varying dimensionality and the effect of mixtures of both active and passive Brownian particles. These models focus on particles in finite space, e.g. with periodic boundary conditions \cite{Sansa2018Velocity,Adam2014Cooperative,Mognetti2013Living,Stenhammar2014Phase,Stenhammar2015Activity,Kudrolli2008Swarming}. However, experimental analogues are rare with the few examples including self propelled robots, colloid systems, and vibrated granular systems \cite{Mognetti2013Living,Narayan2007Long,Julien2010Collective,Giomi2013Swarming}. To our knowledge few corresponding examples exist
 in living systems, although active phase separation has been seen in bacteria \cite{Liu2019Self} and mussels \cite{Liu11905}.
\\\\
Another emerging strand of literature has begun to focus on the role of inertia in self-propelled particle systems, leading to an equation of motion that is second order in time. The presence of inertia has lead to observations of inertial delay between particle velocity and body axis \cite{Scholz2018Inertial}. If the inertial effect is strong enough
it appears that the onset of MIPS occurs at higher P\'{e}clet numbers (a dimensionless ratio of self propulsion speed to the rate of diffusion \cite{RevModPhys.88.045006}) and can vanish for large enough particle mass. Furthermore, before the onset of MIPS, a novel phase coexistence between
``hot'' and ``cold'' regions develops where the kinetic energy per particle (kinetic temperature) is low in the dense phase and high in the dilute phase (a difference of a factor of 100 has been predicted) \cite{Hartmut2020InertialEffects,mandal2019motility}. Similarly it has been found that the presence of inertia drastically changes the dynamics of a rotating micro swimmer \cite{Yuanjian2020Rototaxis}.
Experimentally the realisation of inertia in self-propelled particle systems is seen in active granule systems such as macroscopic ``Bristle bots'' or ``Vibro-bots'' which utilise either a small vibrating motor or are placed on a vibrating plate with angled feet to provide self propulsion \cite{Giomi2013Swarming,Deblais2018Boundariess}. Beyond this, experimental realisations of inertial delay are rare, particularly in living systems.
\\\\
We study experimental footage of whirligig beetles \textit{D. discolor} containing between $N=50$ and $200$ individuals that are moving freely on a water surface within a large circular arena. Figure \ref{fig:Panel1}$a$ depicts the experimental setup with an overlay representing the topological method we employ for local density estimation. 
These water beetles are ellipsoidal in nature with an aspect ratio (from a top down perspective) measured as approximately 2:1 and a body
length of approximately $12\pm1$mm \cite{ferkinhoff1983key}. Whirligig Beetles are a particularly useful study organism due to their lack of
group hierarchy and strong similarity between individuals (both particularly important traits in the context of MIPS). It is also relatively simple
to collect top-down 2D video footage of their movement.
\\\\
Previous studies have focused on their natural
behaviour and include observations of large scale clusters (``rafts''), which form during
the day and can number from $100$s to $1000$s of individuals \cite{Bernd1980Behavioural,Vulinec1989Aggregation}. These structures are noted
for their rapid dispersal (flash expansion events) and reformation when threatened by predators such as fish. The rapid break-up is thought to be caused by a cascading signalling process in which beetles sense (via vision or sensation of water disturbance) the movement of neighbours and react accordingly by moving rapidly and often randomly, with the
onset dependent on the number of visibly startled beetles \cite{Romey2015A}. In particular,
the movement is directed away from the group's geometric centre and not the point of highest density or from the location of the original beetle to startle \cite{Romey2015B}. Here we neglect any possible role of capillary interactions between individual beetles \cite{voise2011capillary}, noting that these are probably less significant for strongly self-propelled particles. Other studies have focused on the individual movement capabilities of Whirligig beetles, with applications
to the design of efficient ``fast'' bio-inspired artificial swimming robots \cite{Philamore2015Row,Jia2015Energy,Bokeon2017Design}. Individual beetles have been previously observed moving with maximum speeds of $160$ body lengths per second (in bursting events), reaching accelerations of $2.86g$, and maximum turning rates of $77.3 \enspace rad/s$ \cite{Fish2003turning,Xu2012Experimental,Voise2009Management}.
\\\\
We report evidence for a density dependent swimming speed which decays as a power law of density for different populations sizes studied. This is indicative of a marked propulsion speed difference between the dense clustered phase and dilute phases that are observed to coexist. We present a simple model, based on the motion of active brownian particles (ABPs), that is able to capture the empirical density probability density function (PDF) observed in the data. This model, which we call Corralled Active Brownian Particles, generates the turning of particles back towards the geometric centre of the cluster. The turning is proportional to a strength coefficient $\tau$ and a power $\alpha \geq 0$ of density.
Finally, we demonstrate the presence of inertial effects in the form of a short inertial delay between the beetle's body axis and its velocity vector.
\section{Results}
\subsection{Speed and Density}
Using individual beetle trajectory data we extract the speed $v$ and density $\rho$ averaged over particles and time. The speed here is defined in the short time limit as the particle displacement between individual frames and is calculated using a central difference method for higher order accuracy.  Note that the crossover to diffusive behaviour is on much longer timescales of 10-30 frames (see SI figure S6, consistent with S5). The density is a local (topological) 
measure of individual beetle density, see methods. Figure \ref{fig:Panel1}$b$ shows the averaged speed for particles with density $\rho$, written $v(\rho)$. This exhibits an empirical power law scaling across a broad regime of densities and appears to be quite consistent across different population sizes. The slope associated with an exponent of $-0.4$ is shown in figure \ref{fig:Panel1}$b$ as a guide to the eye. At the very highest local densities we observe a marked break from the power law to movement speeds increasing with density. We speculate that this may be associated with the coordinated motions of high density domains in the cluster, moving as a rigid body.

\subsection{Orientation-Velocity Correlation}
As shown in figures \ref{fig:Panel1}$c$ and \ref{fig:Panel1}$d$ we find the  body axis leads the velocity by a positive lag time. The correlation function is given by
\begin{align}
    C(\Delta t) = \langle \hat{\boldsymbol{V}}_{i}(t+\Delta t)\cdot \hat{\boldsymbol{n}}(t)\rangle_{t,i} \label{eq:ovcorr}
\end{align}
and measures the average scalar product between the orientation at time $t$ and the instantaneous velocity direction at time $t+\Delta t$ where $\Delta t$ is the time lag. All times are discrete in units of the video frame interval ($\frac{1}{30}$ seconds), hence the discrete points on figure \ref{fig:Panel1}$d$. The average $\langle \dots \rangle_{t,i}$ is over time $t$ and over the beetle trajectory index $i$. We only analyse trajectories that we consider to be reliably collision-free, that is trajectories in which the minimum inter-particle distance over the entire trajectory is greater than a threshold of one body length, (see SI section S1 for details). The positive peak location corresponds to an inertial delay of approximately $13$ ms.
\subsection{Model: Corralled Active Brownian Particles (CABPs)}\label{CABP}
\begin{figure}[ht!]
\centering
\includegraphics[width=0.99\linewidth]{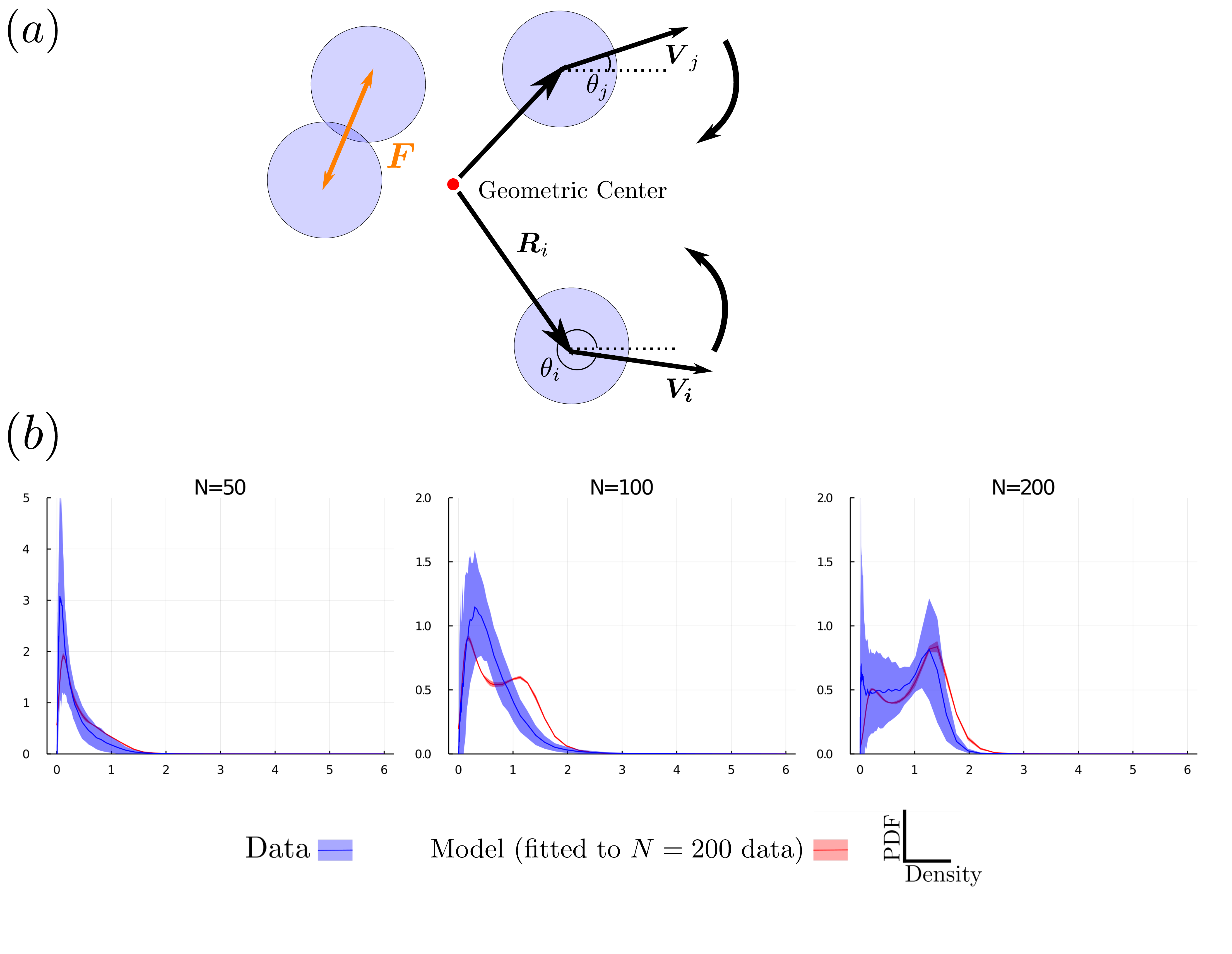}
\caption{$(a)$ Cartoon of the model. The particles are treated as soft, self-propelled disks that repel with a force $F$ when overlapped, as in a standard ABP model. However, they also experience a density dependent re-orientation towards the centre of mass of the swarm (annotated Geometric Center), represented by the curved arrows. The reorientation depends on the local density and is assumed strongest at low densities.
$(b)$ The density PDFs for the experimental data (blue) and the fitted model (red), evaluated for different numbers of particles $N$, as shown. The model dynamics correctly recover coexisting dilute and dense phases for $N=200$ and a dilute phase alone for $N=50$, the phase boundary is around $N=100$.  The model is parameterised once only,  by fitting to the PDF for the $N=200$ human tracked dataset; the red curves for $N=50,100$ are the results of this model evaluated with different particle counts. The solid lines are the mean density distributions (kernel density
estimates), the error bands indicate one standard deviation. Other simulation parameters in this and subsequent figures are $\mu k = 316.2$, $D_{r}=2.34$ rad$^{2}s^{-1}$, and $v_{0}=13.19$ body lengths per second.}
\label{fig:Panel2}
\end{figure}

To model the behaviour of the swarm we develop a minimal particle-based simulation and fit this to the data.
This is based on a  standard ABP model 
(neglecting hydrodynamic interactions and inertia) within the same framework as \cite{Fily2012Athermal} but with
an additional reorientation term included in the orientational dynamics to account for the fact that our system is not contained by periodic boundaries and therefore needs a mechanism to corral the particles into the same region in space - behaviour that is clearly exhibited by the beetles themselves. To this end we introduce a torque that tends to re-orientate particles back towards the centre of mass of the cluster, see figure \ref{fig:Panel2}$a$.
\\\\
This re-orientation, acts separately on all particles and is assigned a strength that depends on the local density. Models that incorporate similar torques, but designed to promote co-alignment, have been used to study the affect of particle alignment on the onset of MIPS \cite{Sansa2018Velocity}. Torque terms can also arise naturally as a consequence of particle geometry \cite{Damme2019Interparticle} although we neglect this in our model.
For simplicity our model employs self-propelling polar disks of uniform radius $a$ (with the 
diameter $2a$ considered equal to one beetle length), self-propelled along the body axis (polarity) $\hat{\boldsymbol{n}}_{i} = [\cos\theta_{i},\sin\theta_{i}]^{T}$, of the $i$th particle. Propulsion along the $\hat{\boldsymbol{n}}_{i}$ direction involves a constant magnitude speed $v_{0}$. The collisions are soft-body interactions with a harmonic force on particle $i$ due to $j$ of $\boldsymbol{F}_{ij} = -k(2a-r_{ij})\hat{\boldsymbol{r}}_{ij}$ for $r_{ij}<2a$, $\boldsymbol{F}_{ij}=0$ otherwise. This involves the interparticle separation vector $\boldsymbol{r}_{ij} = \boldsymbol{r}_{j}-\boldsymbol{r}_{i}$, with $\boldsymbol{r}_{i}$ the position of particle $i$, its magnitude $r_{ij}$ and where a hat ($\>\hat{}\>$) denotes a unit vector throughout. The particles follow over-damped Langevin equations of motion given by
\begin{align}
        \partial_{t}\boldsymbol{r}_{i} &= v_{0}\hat{\boldsymbol{n}}_{i} + \mu \sum_{j\neq i} \boldsymbol{F}_{ij} \label{eq:trans} \\
        \partial_{t}\theta_{i} &= \eta_{i}(t) + \tau\rho_{i}(t)^{-\alpha}(\hat{\boldsymbol{R}}_{i}\times\hat{\boldsymbol{V}}_{i})\cdot\hat{\bf z} \label{eq:rot}
\end{align}
Here $\mu$ is a mobility parameter. However, both this and the elastic constant $k$ only appear in the product $\mu k$ and so this does not introduce an additional control parameter. The angular noise $\eta_{i}(t)$ has zero mean and is Gaussian distributed according to $\langle\eta_{i}(t)\eta_{j}(t')\rangle = 2D_{r}\delta_{ij}\delta(t-t')$ where $D_{r}$ is a rotational diffusion coefficient; a positional noise term can be neglected here. The  re-orientation term in the angular dynamics generates reorientation towards the centre
of mass with $\boldsymbol{R}_{i} = \langle \boldsymbol{r}_{j}\rangle - \boldsymbol{r}_{i}$ the vector pointing from a particle to the centre of mass, and
$\boldsymbol{V}_{i}$ the instantaneous velocity of particle $i$. The dot product with $\hat{\bf z}$ (out of the plane of motion) converts this to a signed scalar, positive for anticlockwise turns and negative for clockwise. Finally the factor $\tau\rho_{i}^{-\alpha}$ (with $\alpha > 0$) represents a simple choice for the density dependence of the reorientation. All simulations are carried out in unbounded 2D space, with the density that emerges within the cluster controlled by the values of $\tau$ and $\alpha$. All quantities are reported as dimensionless throughout with times scaled in seconds and lengths scaled with the beetle particle length (disk diameter).
\\\\
To reduce the dimensionality of the fitting process we determine the value of the rotational diffusion coefficient by directly fitting to the mean square angular displacement of the $N=200$ beetle data, yielding $D_{r}=2.34$ $rad\> s^{-1}$. We also fix the self propulsion speed $v_{0}$ as the average speed of beetles that are freely moving (see SI for a discussion of collision free trajectories), yielding $v_{0}=13.19$ body lengths per second. We fit the free parameters ($\mu k,\alpha, \tau$) by minimising the error, using a Bayesian optimisation technique (see Methods), between the empirical density distribution (PDF) for the $N=200$ human tracked beetle data (only) and the density distribution obtained from a simulation with $N=200$ particles (see the methods section for details). The resultant density distributions, for {\it all} data sets are shown in figure \ref{fig:Panel2}$b$, together with the results from simulations of the model (fitted to $N=200$ only, not re-parameterised for each dataset) containing the corresponding number of particles $N=50, 100, 200$.  Best-fit parameter values are shown in Table~\ref{tab:fit}. We include an example
simulation for the fitted model evaluated with $N=200$ particles as
SI movie S7.
\begin{table}[ht!]
    \centering
    \begin{tabular}{|c|c|c|c|c|c|}
    \hline
        \text{ Parameter} & $\alpha$ & $\tau$ $(s^{-1})$ & $\mu k$ $(s^{-1})$ & $v_{0} \enspace (s^{-1})$ & $D_{r} \enspace (\text{rad}^{2}s^{-1})$ \\
    \hline
        \text{ Best fit value } & 1.1 & 19.6 & 316 & 13.19 & 2.34 \\
    \hline
    \end{tabular}
    \caption{Best-fit values of the control parameters. These are inferred by fitting the results of simulations performed on $N=200$ particles, using equations of motion \ref{eq:trans} and \ref{eq:rot}, to the $N=200$ beetle dataset. The fit metric is a least-squares measure of the density PDF. We include $v_{0}$ and $D_{r}$ in this table but note that they are fitted
    to data \textit{before} using Bayesian optimisation to find $\tau$, $\alpha$, and $\mu k$ to reduce dimensionality. We measure all lengths in units of the mean beetle body length.}
    \label{tab:fit}
\end{table}
Also shown in figure \ref{fig:Panel2}$b$ are the results of the best-fit model, simulated at different $N$ values, as shown. As the number of particles increases we
observe a clear transition from a uni-modal density distribution, with a single peak at low density (similar to a dilute gas phase) to a bi-modal distribution (similar to a dilute gas coexisting with a dense liquid). A similar trend has been observed in simulations \cite{Fily2012Athermal}, 
where $N$ directly controls the density and phase separation (bi-modality) appears only above a critical threshold.
\subsection{Phase diagram in $\alpha$ -- $\tau$ space}
\begin{figure}[ht!]
    \centering
    \includegraphics[width=0.99\textwidth]{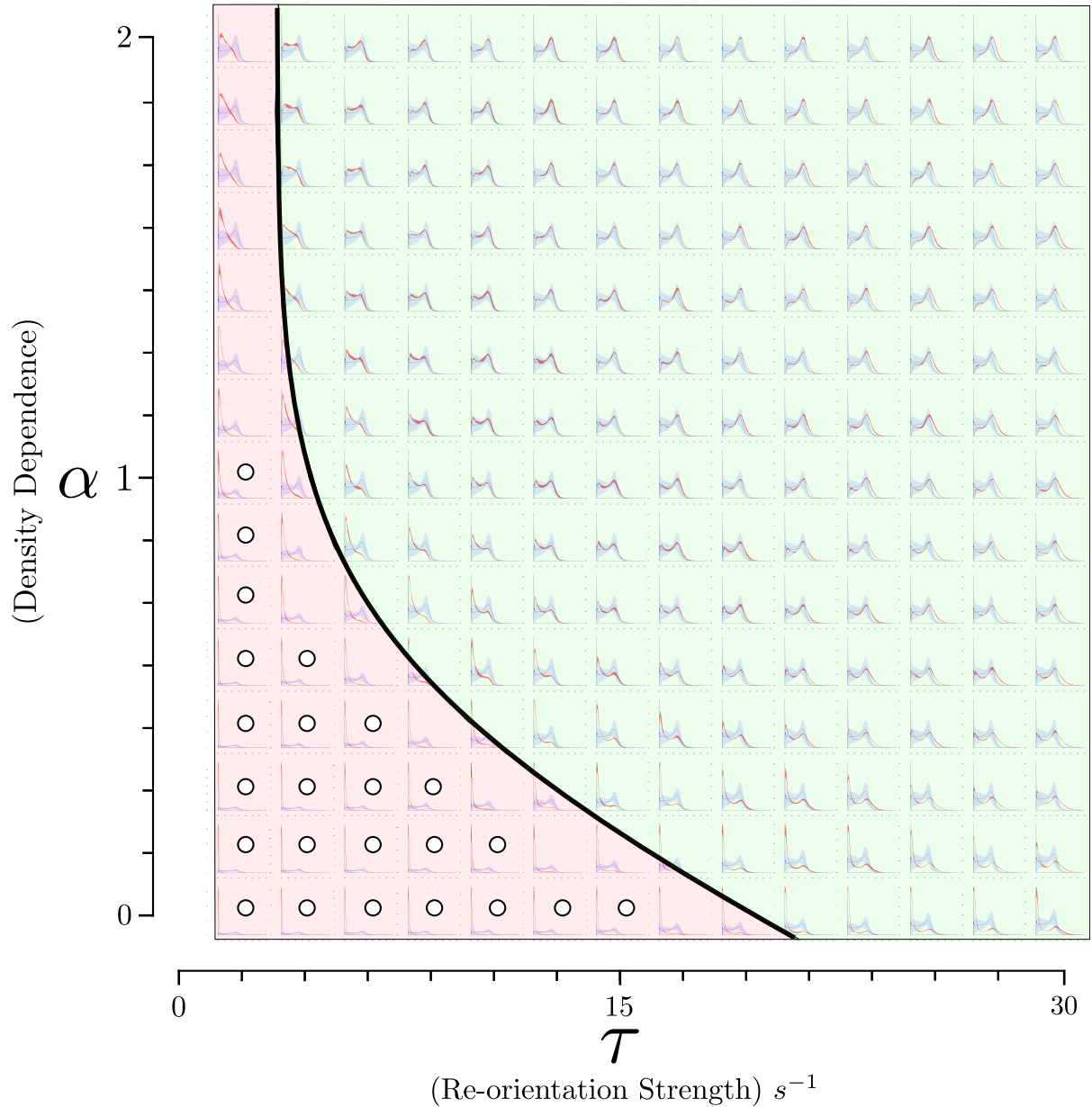}
    \caption{Phase diagram for our CABP model in $\alpha$--$\tau$ space, with the one phase (dilute gas)  highlighted in red and the two phase (gas and dense liquid) in green. Each of the $14\times15$ sub-panels (the $\tau=0$ column is excluded) shows the density PDF obtained from a dynamical simulation for $N=200$ particles at the corresponding $\tau$, $\alpha$ values (red curve, with SD error band). Also shown on each panel is the density PDF from the experimental data at the same $N$ value (blue curve, with SD error band, the same on each panel). The best fit parameters from Table~\ref{tab:fit} correspond roughly to the system shown in row 7, column 9, deep in the two phase region. The approximate location of the phase boundary denoting the appearance of a high density phase is identified by eye. Circles overlaid on the plots indicate that we are
    not confident that the density has reached an equilibrium state  for these systems within 30 s real time. This means that the gas phase density may be even smaller than shown; the phase boundary is unaffected. The equilibration time used was between 10 s and 30 s, longer times being required as we move towards the bottom left corner, where re-orientation is weak.}
    \label{fig:phase-diagram}
\end{figure}

Using the fitted value of $\mu k$, and the values of $v_{0}$ and $D_{r}$ previously extracted directly from the data, we compute the density PDF generated by dynamical simulation of the model over a large space of reorientation functions, parameterised by different values of $\alpha$ and $\tau$ that control the functional form of the reorientation, as defined in Eq~\ref{eq:rot}. The results are displayed in figure \ref{fig:phase-diagram} where each of the 210
sub-panels represents a density PDF like those shown in figure \ref{fig:Panel2}$b$. The density PDF for the $N=200$ beetle data is shown, identically, on every sub panel (blue). The PDF obtained by simulating each model, parameterised with different $\alpha$ and $\tau$ values, differ between sub-panels (red).
\\\\
The prefactor to the reorientation strength $\tau$ increases across the rows, from left to right. This means that the systems represented in panels on the left of a row have an unambiguously weaker turning reorientation than those on the right. Weaker reorientation leads to a more diffuse cloud of particles and a lower mean density. This is consistent with the one-phase region of dilute gas being located on the left hand side of (rows of) this diagram. Moving down the columns corresponds to reorientation strength having a stronger density dependence (exponent) $\alpha$. Since the dense liquid phase has a dimensionless density $1\lesssim\rho\lesssim 2$ there is relatively little variation of the reorientation strength in the liquid phase moving down the columns but the gas phase typically has a density $\rho\ll 1$ and there is a correspondingly stronger influence on reorientation in this phase.
\\\\
In figure \ref{fig:phase-diagram} we also identify the approximate location of the phase boundary between a one-phase gas (low density, uni-modal PDF) and the two phase gas-liquid co-existence (bi-modal PDF), see SI for similar phase diagrams for $N=50$ and $N=100$. The slow equilibration of systems with parameter values in the bottom left corner of figure \ref{fig:phase-diagram}, corresponding to weak reorientation, does not affect our conclusions: the density PDFs in these systems may still be slowly shifting to even {\it smaller} densities, meaning that they are even deeper in the gas phase than they appear. It is significant that, in all cases, MIPS-like phase coexistence arises for appropriate parameters $\tau$ and $\alpha$, corresponding to sufficiently strong turning, or ``corralling'', effect. This is sufficient to maintain high enough overall system densities in much the same way that ABP models need to sit above some critical density for MIPS to arise \cite{Fily2012Athermal}.

\section{Discussion}
While there is enormous contemporary interest in the physics of active particle systems, it remains unclear how broadly these ideas can be tested experimentally, especially in living analogues. We focus on three separate concepts.
\\\\
Phenomenological velocity-density relations play a central role in foundational studies of MIPS \cite{Cates2015Motility} and can be viewed as a fundamental feature in our current understanding of the phenomenon. However, we are unaware of studies that directly analyse this relationship, with Liu \textit{et al} \cite{Liu11905} a notable exception. In the present work we report a particularly simple power-law scaling form for this relationship in whirligig beetles that seems to hold over two orders of magnitude in density. This may motivate the development of active field theories that directly encode such a power law.
\\\\
Secondly, there has been significant recent interest in the role of inertial effects in collections of ABPs \cite{Scholz2018Inertial,Hartmut2020InertialEffects}. We present evidence for the existence of such inertial delay in macroscopic living systems, with inertial delays in the ten millisecond regime. This inertial delay can be described as ``short'' given that the distance moved in body lengths and the root mean squared angular diffusion are both only $O(10^{-1})$ on this timescale. This gives us some reassurance that non-inertial models, such as employed in section \ref{CABP}, may be adequate at the semi-quantitative level.
\\\\
Thirdly, we show that a condensed ``liquid-like'' phase of whirligigs can coexist with a dilute ``gas-like'' phase. This is highly reminiscent of MIPS \cite{Fily2012Athermal}, a phenomenon that has been widely studied but lacks experimental realisations, particularly in living analogues. In order to study this phenomenon we developed a model of corralled ABPs that turn inwards, providing a mechanism to control the density of the swarm in open space that seems broadly plausible in its mechanism. We fit this model to experimental data and obtain results that reproduce the bimodal density PDF and MIPS-like coexistence, provided the systems are large enough. Finally, we speculate that our model could be probed experimentally by studying a group of Whirligig beetles ``doped'' with robotic beetles programmed to either turn towards the geometric centre, as here, or responding to other interactions, such as nearest neighbour alignment or attraction/repulsion.
\\\\
Our work provides a new way of understanding the behaviour of this insect, and the mechanism for the formation of dense clusters of individuals, in terms of the MIPS paradigm. MIPS is a phenomenon that has recently been identified in the context of non-equilibrium physics. In this literature self-propelled particles (ABPs) with purely repulsive contact forces have been shown to phase separate, forming similar clusters. The fact that clustering occurs in the absence of any attraction is a signature of the out-of-equilibrium nature of the particles, i.e. that they are motile. By analogy we suggest that the phenomenon of beetle clustering need not involve any direct attractive interactions. This was far from obvious at the outset. We have shown that the beetle’s behaviour is consistent with a CABP model that we develop. This reproduces MIPS-like clustering. We do not view the corralling (turning) in our CABP model as a form of attraction but believe that it is better viewed as providing a global constraint on the density, necessary in unbounded space. This is because it involves no pairwise attractive interactions. Also, in classical ABP systems, density is regulated by the walls of a box (or its periodicity), limiting its volume, but one wouldn’t say this confinement provides an ``attraction’’, rather that it serves to fix the density. The identification of a model insect system that exhibits MIPS-like clustering is also likely to be of keen interest to Physicists, both as a rare example in multicellular organisms but also for what it tells us about empirical velocity-density relationships in such systems.
\section{Methods}
\subsection{Experimental Data} \label{sec:data}
Our raw data consists of footage from experiments on varying population sizes ($50$,$100$,$200$) of whirligig beetles (\textit{D. discolor} collected from the Raquette River in Potsdam, NY, USA) in a cylindrical tank of water with a diameter of $1$ m and individual beetles measuring $12\pm 1$ mm in length along the major axis. Each video was filmed for
a period of around $5$ minutes at a resolution of $1920\times 1080$ pixels and a frame rate of $30$ frames per second (see SI movie S6 for a high contrast example at $N=200$). The camera was situated 1.96 m above the water level, a fluorescent light 
illuminated the apparatus at 730 Lux. Beetles where not startled or given time-varying external stimulus
during filming and were allowed 20 minuets acclimatisation time before filming. Beetles were fed at 7:00 am and 7:00pm (at the time of filming sunrise and sunset where measured at 5:30-6:00am and 8:30-9:00pm each day) and experiments performed between these times.
For the $N=200$ population,
footage was tracked by hand, and for each population footage was tracked
using a modified version of the network flow formulation (see an outline in our methods, section \ref{sec:tracking}) which was validated
by reference to the human tracked set of data. These highly accurate tracks for individual beetles represent their coordinates (in the lab frame) at each time point, together with their orientations,
velocity, and density which we use in our analysis of the dynamics of the collective.
In the experimental footage we observe a pronounced change in behaviour across the three swarm sizes. 
The footage covering the $50$ beetles population is more erratic in character with mostly short lived cluster
formation and a generally elevated propulsion speed (see SI figure S2 for speed distributions). The $N=100$ and $200$ beetles swarm around relatively stable clusters. In these systems, a fraction of individuals reside in a more dilute ``corona'' around the main cluster, corresponding to the dense and dilute phases respectively. The dense regions are notable for significantly decreased self-propulsion speed.
\subsection{Tracking}\label{sec:tracking}
The individual beetle tracks were generated from centre of mass coordinates and (major axis) orientations for each beetle using a modification of the network flow method for multi-object tracking \cite{Zhang2008Global}. Our method \cite{trackingdraft}
takes advantage of the fact we can assume conservation of beetle population, whereas Li Zhang \textit{et al} track pedestrian data in which many individuals leave and enter the video frame. The coordinates and orientations were gathered from raw video footage by training a neural network on a set of validation footage which was tracked by hand (logging positions orientations and linking beetle tracks). Human marked tracks ($N=200$ data) were also used to validate the tracking method, and to fit the model.
\subsection{Density} \label{sec:dense}
To calculate the local density at a beetle's centre of mass, we use a method based upon the Delaunay triangulation to assign an area and therefore a number density to each beetle. Delaunay triangulation based methods of interpolating density fields from point data have been used successfully in astronomy \cite{schaap2007delaunay}, and the estimation of group density has been computed using alpha shapes to calculate an area fraction \cite{Sosna20556}, which is closely related to the Delaunay triangulation.\\\\Our method identifies the number density $\rho(\boldsymbol{r}_{i})$ of a point $\boldsymbol{r}_{i}$ as equation \ref{eq:density}. For notation, we write the Delaunay triangles with common vertex, $\boldsymbol{r}_{i}$, as $T_{i}^{(j)}$ so that $j$ is an index over this set of triangles. Note that points on the edge of the convex hull, will generally be associated with fewer
Delaunay triangles since the Delaunay triangulation only tessellates the convex hull. The area of $T_{i}^{(j)}$ is written as $A_{i}^{(j)}$, and similarly the angle made at point $i$ by the two edges of the triangle $T_{i}^{(j)}$ ending at $i$
is $\theta_{i}^{(j)}$. Refer to figure \ref{fig:Panel1}$a$ for a visual representation of these quantities on an example Delaunay Tessellation computed from data. Using this notation the density at point $i$ is equation \ref{eq:density}
\begin{align}
    \rho(\boldsymbol{r}_{i}) = \frac{1}{2}\frac{\sum_{j}\theta_{i}^{(j)}}{\sum_{j} \theta_{i}^{(j)}A_{i}^{(j)}} \label{eq:density}
\end{align}
To derive equation \ref{eq:density}
we consider particle $i$ as contributing $\theta_{i}^{(j)}$ of it's ``mass" to the area $A_{i}^{(j)}$, this
gives a normalised ``area per particle" of $\frac{1}{\pi}A_{i}^{(j)}$ for triangle $T_{i}^{(j)}$. Then to calculate a normalised density we take the average (weighted by the angles $\theta_{i}^{(j)}$) of this normalised area per particle and invert it yielding $\pi\sum_{j}\theta_{i}^{(j)}/(\sum_{j} \theta_{i}^{(j)}A_{i}^{(j)})$. Since a point in the interior
will satisfy $\sum_{j}\theta_{i}^{(j)} = 2\pi$ and the boundary points will satisfy $\sum_{j}\theta_{i}^{(j)} < 2\pi$ we divide this expression by $2\pi$ to arrive at equation \ref{eq:density}.
\\\\
The advantage of this approach, and our key reason for taking it, is that it avoids dividing by areas individually. This can lead to arbitrarily large local densities in
the case of one triangle having an extremely small area (``sliver triangles''). These small triangles can form when three points in a Delaunay triangulation are arbitrarily close to being collinear.
\subsection{Model Fitting} \label{sec:fitting}
Our model parameters were tuned using a Bayesian optimisation
framework \cite{bayes-opt}. The quality of fit is reported by the mean square error between the simulation density distribution and the empirical distribution obtained from the data.
The fitting itself employs a Gaussian kernel density estimator on the experimental and simulated density distributions
with bandwidth parameter chosen according to Silverman's rule \cite{silverman1986density}.
To calculate the fitting error the mean square error between the two kernel density estimates
was computed over a the range of the experimental data.\\\\
In all our simulations we take a time step of $\frac{1}{900}$ seconds. We discard the initial $4,500$ time steps to eliminate short lived initial transients in the global density. Initial conditions for the simulations are randomly drawn from uniform distributions, both for
position and orientation, with positions restricted to an square region of size corresponding to a density $\rho_{\rm init}=0.5$. Results are reported as averages over three simulations with different random initial conditions, each with the same set of fitted parameters.
\subsubsection{Optimising Runtime}
Running numerous simulations of the CABP model at $N=200$ as is required for the fitting process is computationally expensive. In order to most efficiently deploy computation resources we take advantage of large scale parallelisation on GPUs. We implement the solver for the active Brownian particle model as a GPU (CUDA \cite{Nickolls2008}) algorithm, parallelising across particle index.
\subsection{Data and Code Availability}
All data and the code used for its analysis are available at the GitHub repository \textit{https://github.com/harveydevereux/CUDA-Whirligigs}. Instructions are provided in the top level README.md
\subsection*{Authors' Contributions}
M.S.T, S.T, and H.L.D designed research;
C.R.T and S.T obtained data; H.L.D wrote code for the model and analysis, and performed simulations; M.S.T, S.T, and H.L.D analysed the data; M.S.T, S.T, C.R.T and H.L.D wrote paper, and gave final approval.
\subsection*{Competing Interests}
We declare no competing interests.
\subsection{Funding}
\small{
Funding for this work was provided by UK Engineering and Physical Sciences Research Council though the Mathematics for Real-World Systems Centre for Doctoral Training Grant EP/L015374/1 (HLD). Facilities for all
numerical work were provided by the Scientific Computing Research Technology Platform of the University of Warwick. This material is also based upon work supported by the National Science Foundation Graduate Research Fellowship (to CRT) under Grant No. DGE-0646086. ST acknowledges support from the Department of Atomic Energy, Government of India, under project no. 12-R\&D-TFR-5.04-0800 and 12-R\&D-TFR-5.10-1100, the Simons Foundation (Grant No. 287975) and the Max Planck Society through a Max-Planck-Partner-Group at NCBS-TIFR. MST acknowledges the support of a long term fellowship from the Japan Society for the Promotion of Science, a Leverhulme Trust visiting fellowship and the peerless hospitality of Prof Ryoichi Yamamoto (Kyoto). }
\subsection{ACKNOWLEDGEMENTS}
\small{
We would like to thank Harmut L\"owen (D\"usseldorf) for his discussions on inertial effects in SPPs. We
would also like to thank William Romey, at SUNY Potsdam, for his help collecting the whirligigs, and Noor Alkazwin for her tireless work tracking whirligig beetles by hand.}

\section*{Supplementary Material}
\subsection*{S1: Collision Free Statistics}
\begin{figure}[ht!]
    \centering
    \includegraphics[width=0.75\textwidth]{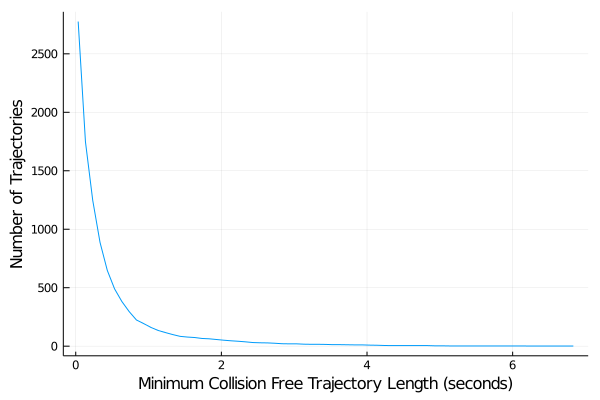}
    \captionsetup{labelformat=empty}
    \caption{Figure S1: A plot of the number of "collision free trajectories" which exceed a time length
    given on the $x$-axis. Most trajectories are less the $1$ seconds long, with a small
    number ranging into $6$ seconds in length. For the delay between the beetle body axis
    orientation and it's velocity direction, the peak is located at an order of $10^{-1}$ seconds.
    Data includes $N=200$ trajectories.}
    \label{fig:cf-count}
\end{figure}

To filter out collisions in the trajectory data for each beetle track we find the Euclidean distances between its position and all others. We consider a sequence of frames to be collision
free if the shortest distance between agents remains strictly greater than one body length for all frames in that trajectory. Mathematically this criterion is given in equation \ref{eq:col}, using
$\theta(x)$ as the Heaviside step function, and $L$ as the body length of a beetle, and
$\boldsymbol{x}_{i}(t)$ is beetle $i$'s position at time step $t$.
\begin{equation}
    i\enspace\text{collision free at time}\enspace t \text{ iff } \sum_{j\neq i} \theta(|\boldsymbol{x}_{i}(t)-\boldsymbol{x}_{j}(t)|_{2}-L) = 0 \label{eq:col}
\end{equation}
Computing this criterion across time reveals any segments of trajectory for any beetle $i$ that are ``collision free'' (Figure S\ref{fig:cf-count}).
\subsection*{S2: Speed Distribution and Group Size}
\begin{figure}[ht!]
    \centering
    \includegraphics[width=0.9\textwidth]{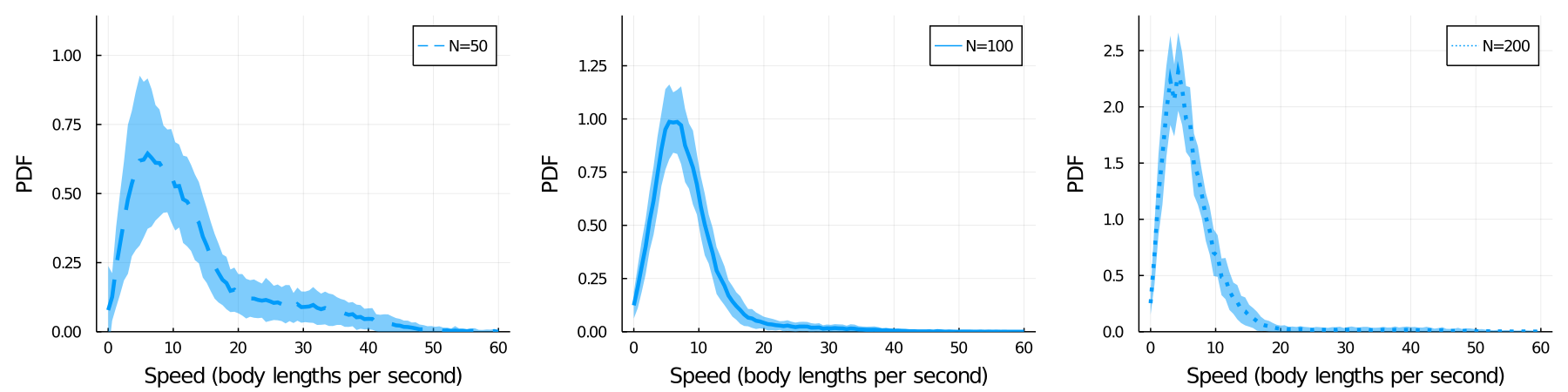}
    \captionsetup{labelformat=empty}
    \caption{Figure S2: Speed distribution for each group size. The error ribbon
    represents one standard deviation between estimates of the probability density function across ($100$) linearly spaced temporal bins.}
    \label{fig:speed-dist}
\end{figure}

For each dataset we measure the speed of each beetle at each time step by
computing the magnitude of the velocity in equation \ref{eq:velocity}, where
$\Delta t = \frac{1}{30}$ is the frame rate of our data and the velocity is
\begin{equation}
    {\bf v}_i(t)= \frac{\boldsymbol{x}_{i}(t+\Delta t) - \boldsymbol{x}_{i}(t-\Delta t)}{2\Delta t} \label{eq:velocity}
\end{equation}
Figure S\ref{fig:speed-dist} shows the empirical probability density function of beetle speed. The error ribbon comes from computing the speed
distribution in $100$ separate batches over the length of the data, the solid line
indicates the mean of these distributions. Similar to the density we
observe a transition from higher speeds to lower ones from large to small
groups.
\subsection*{S3: N=50,100 Phase diagrams}
\begin{figure}[ht!]
    \centering
    \includegraphics[width=\textwidth]{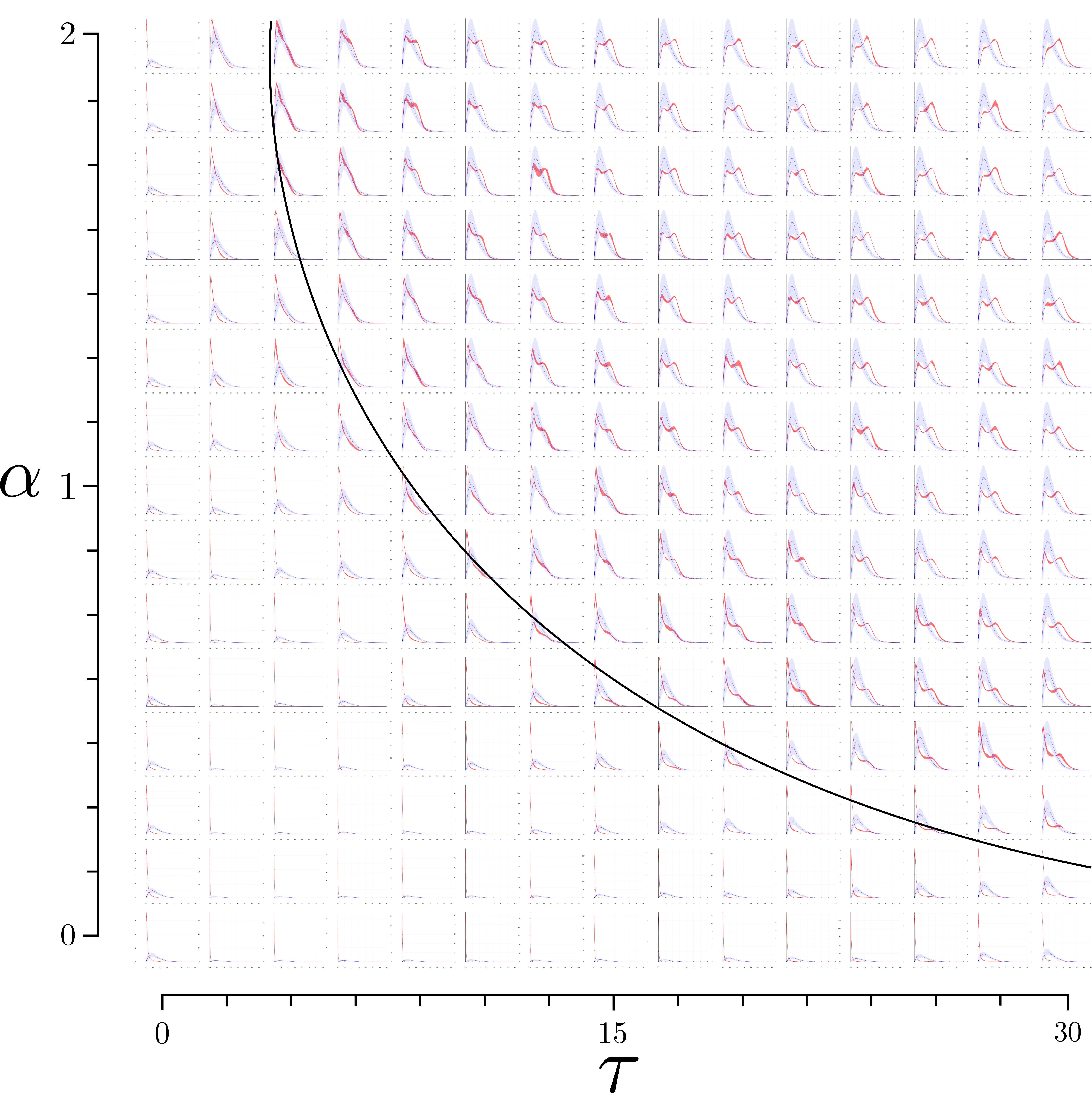}
    \captionsetup{labelformat=empty}
    \caption{Figure S3: Plot of the result of varying $\alpha$ and
    $\tau$ using the model parameters fitted to the $N=200$
    data set, evaluated with $N=100$.}
    \label{fig:100}
\end{figure}

\begin{figure}[ht!]
    \centering
    \includegraphics[width=\textwidth]{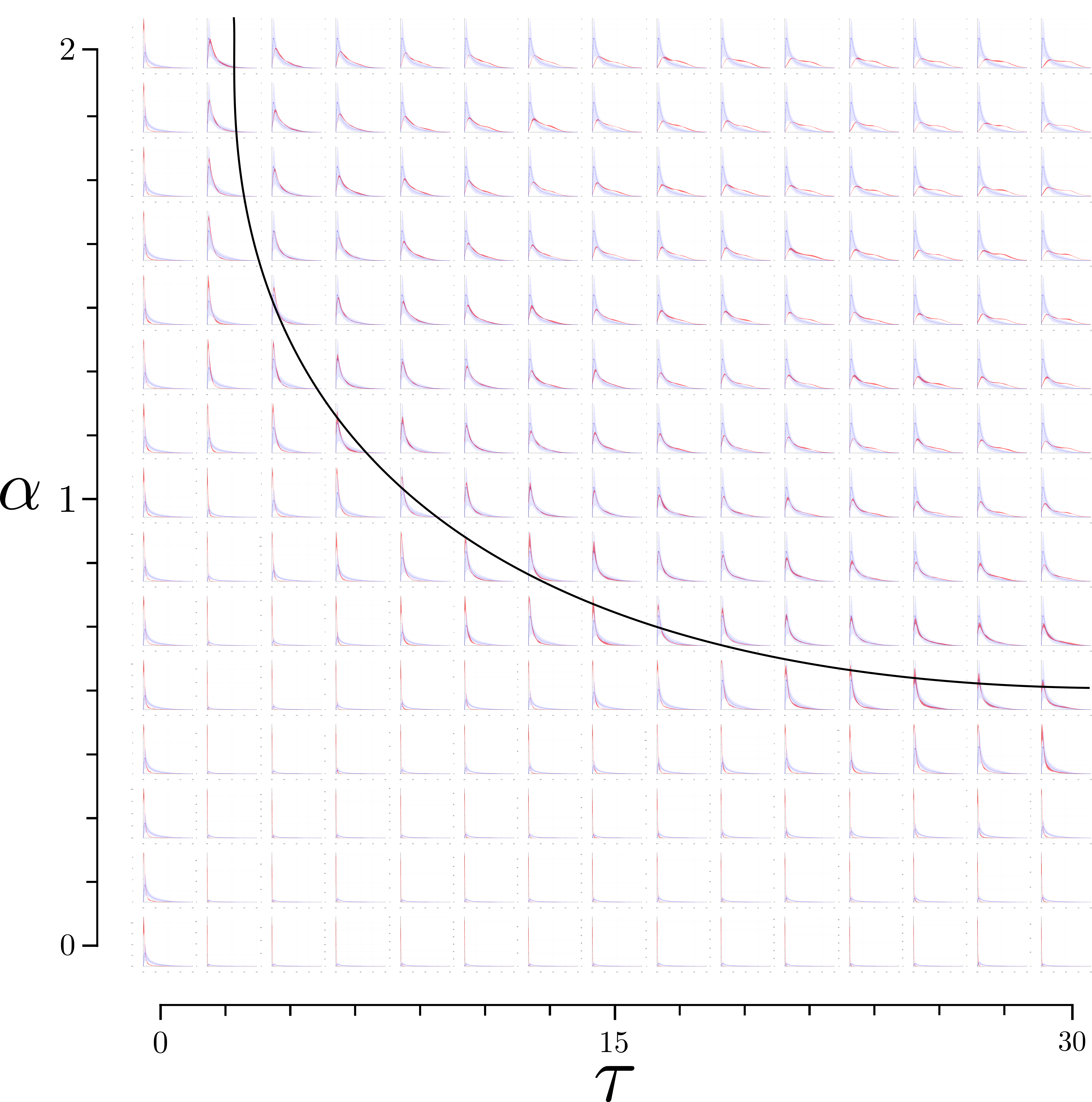}
    \captionsetup{labelformat=empty}
    \caption{Figure S4: Plot of the result of varying $\alpha$ and
    $\tau$ using the model parameters fitted to the $N=200$
    data set, evaluated with $N=50$.}
    \label{fig:50}
\end{figure}

In figures S\ref{fig:100} and S\ref{fig:50} we show the $\alpha$-$\tau$ plane
for population of $N=100$ and $N=50$ compute via the model fitted to the $N=200$
beetle data. We note that in the $N=50$ and $N=100$ cases, good fits to the empirical density distributions exist at different $\alpha$ and $\tau$ values.
\section*{S5: Diffusion}
By computing the mean square angular displacement for discrete time steps
$t$ and discrete time lags $\tau$, we can estimate the diffusion constant
for our equations of motion by finding the gradient of the linear
region and comparing it to $2D_{r}t$. Figure S\ref{fig:angular-diffusion} shows
this analysis for the $200$ group size.
\begin{equation}
    \langle \Delta\theta^2\rangle(\tau) = \frac{1}{NT}\sum_{i=1}^{N}\sum_{t=0}^{T}(\theta_{i}(t+\tau)-\theta_{i}(t))^{2}\label{eq:msad}
\end{equation}
\begin{figure}[ht!]
    \centering
    \includegraphics[width=0.75\textwidth]{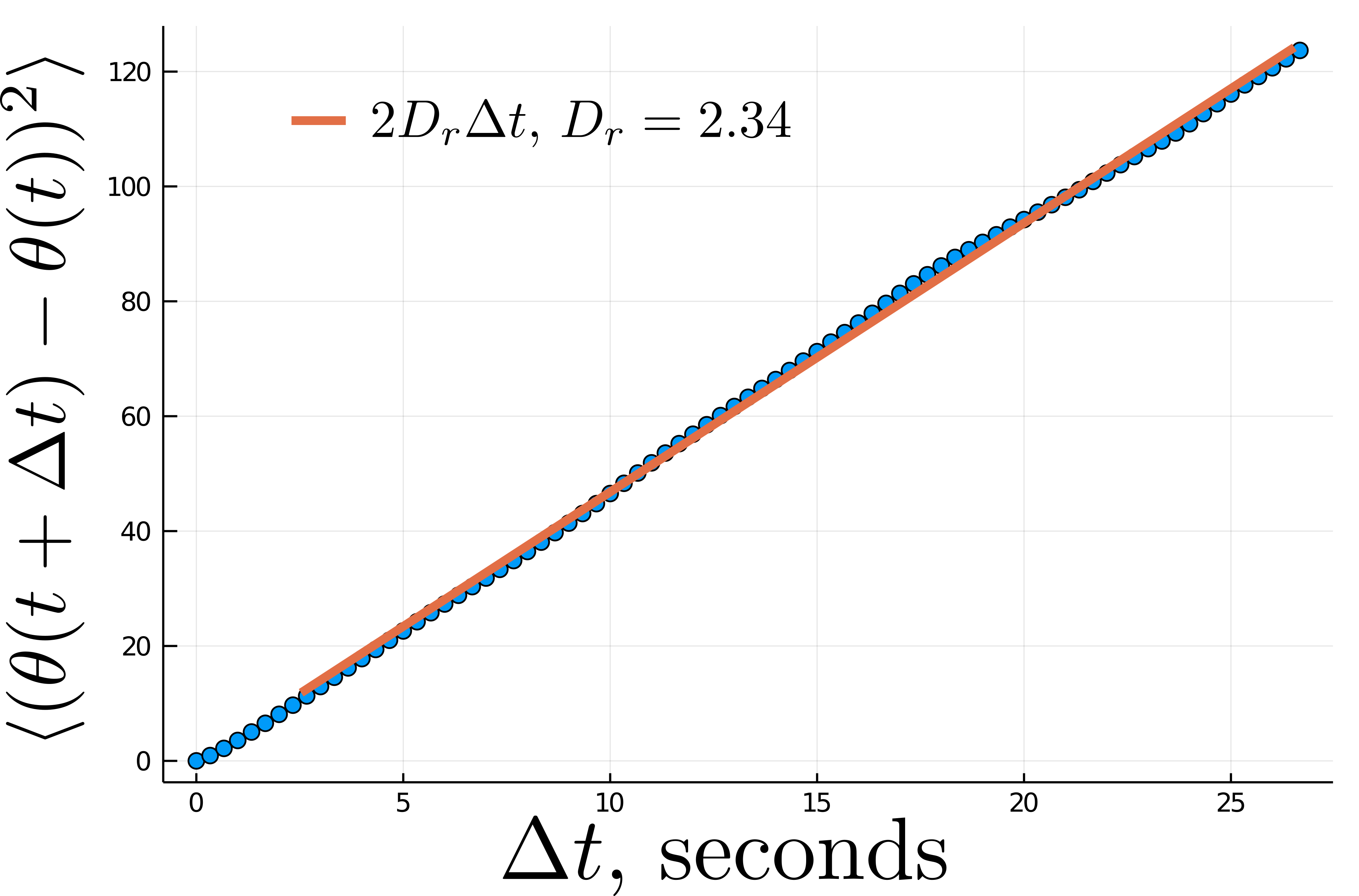}
    \captionsetup{labelformat=empty}
    \caption{Figure S5: Beetles undergo featureless angular diffusion. The average mean square angular displacement after time $\Delta t$ is shown, computed from the $N=200$ 
    dataset. The solid line indicates a linear fit.}
    \label{fig:angular-diffusion}
\end{figure}
\begin{figure}[ht!]
    \centering
    \includegraphics[width=0.75\textwidth]{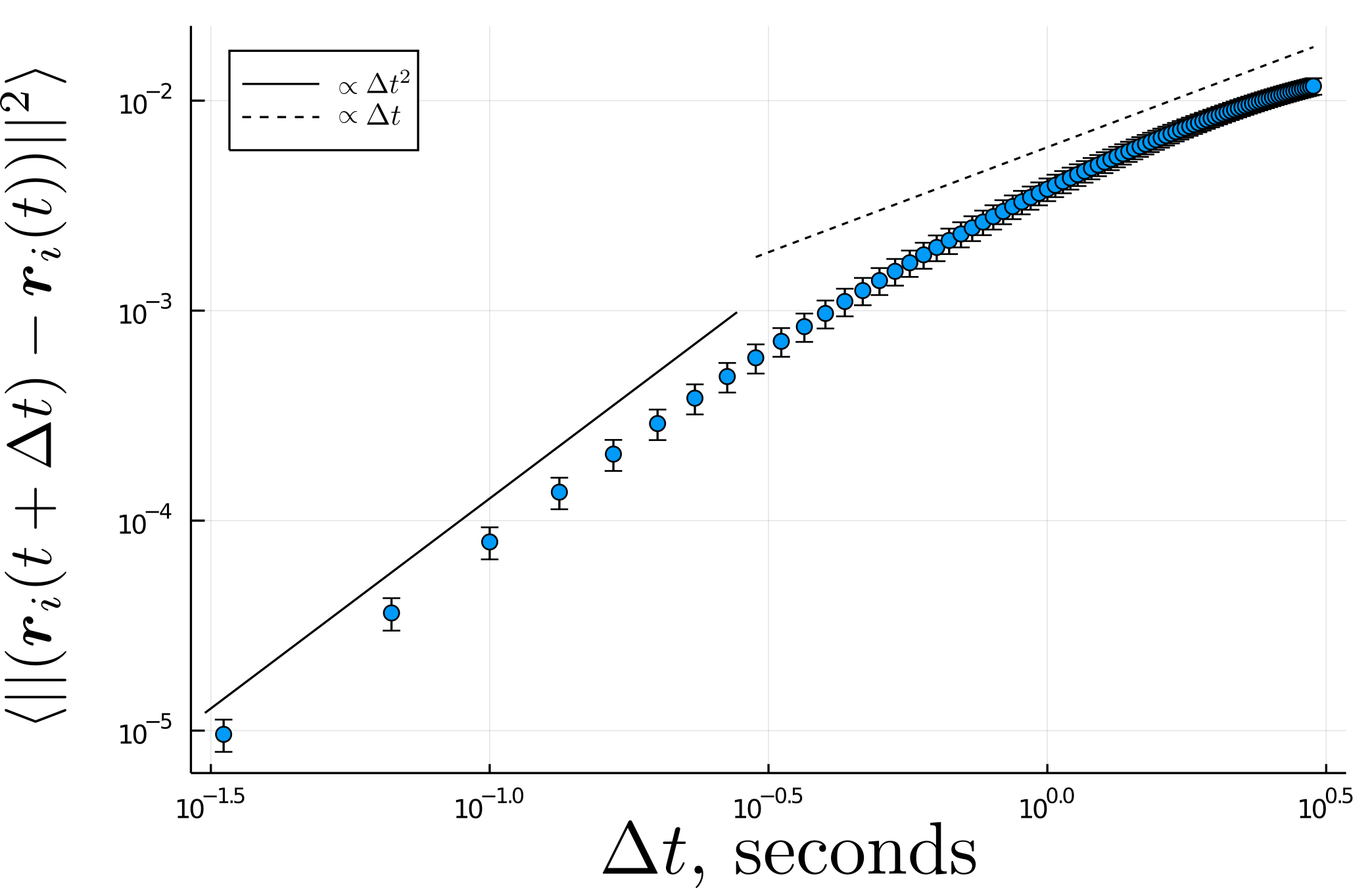}
    \captionsetup{labelformat=empty}
    \caption{Figure S6: Beetles exhibit a clear crossover in which the mean squared displacement crosses over from $\sim \Delta t^2$ (persistent) to $\sim \Delta t$ (diffusive). The average mean square displacement after time $\Delta t$ is shown, computed from the $N=200$ 
    dataset.  Note that we see some signs of an even slower MSD for the longest delay times. This is first due to the swarm clustering within a finite area and, ultimately ,by the fact the beetles are confined to a 1 m diameter arena giving a strict upper bound to their displacement.}
    \label{fig:angular-diffusion}
\end{figure}

\subsection{Movie S7}
Movie 7 shows a typical snapshot
of footage from the experiments. the video has been processed to high contrast to aid in discerning 
individual beetles.
\subsection*{Movie S8}
Movie S8 shows a $30$ s snapshot 
of our fitted CABP model, with 
a population of $N=200$ particles.
Each particle is shown as a blue disc with a red arrow indicating it's orientation. 

\begin{thebibliography}{10}

\bibitem{Fily2012Athermal}
Yaouen Fily and M.~Cristina Marchetti.
\newblock Athermal phase separation of self-propelled particles with no
  alignment.
\newblock {\em Phys. Rev. Lett.}, 108:235702, Jun 2012.

\bibitem{Palacci2013living}
Jeremie Palacci, Stefano Sacanna, Asher~Preska Steinberg, David~J. Pine, and
  Paul~M. Chaikin.
\newblock Living crystals of light-activated colloidal surfers.
\newblock {\em Science}, 339(6122):936--940, 2013.

\bibitem{Buttinonu2013Dynamical}
Ivo Buttinoni, Julian Bialk\'e, Felix K\"ummel, Hartmut L\"owen, Clemens
  Bechinger, and Thomas Speck.
\newblock Dynamical clustering and phase separation in suspensions of
  self-propelled colloidal particles.
\newblock {\em Phys. Rev. Lett.}, 110:238301, Jun 2013.

\bibitem{redner2013a}
Gabriel~S. Redner, Aparna Baskaran, and Michael~F. Hagan.
\newblock Reentrant phase behavior in active colloids with attraction.
\newblock {\em Phys. Rev. E}, 88:012305, Jul 2013.

\bibitem{Cates2008statistical}
J.~Tailleur and M.~E. Cates.
\newblock Statistical mechanics of interacting run-and-tumble bacteria.
\newblock {\em Phys. Rev. Lett.}, 100:218103, May 2008.

\bibitem{Schnitzer1993Theory}
Mark~J. Schnitzer.
\newblock Theory of continuum random walks and application to chemotaxis.
\newblock {\em Phys. Rev. E}, 48:2553--2568, Oct 1993.

\bibitem{Cates2015Motility}
Michael~E. Cates and Julien Tailleur.
\newblock Motility-induced phase separation.
\newblock {\em Annual Review of Condensed Matter Physics}, 6(1):219--244, 2015.

\bibitem{Sansa2018Velocity}
E.~Ses{\'{e}}-Sansa, I.~Pagonabarraga, and D.~Levis.
\newblock Velocity alignment promotes motility-induced phase separation.
\newblock {\em {EPL} (Europhysics Letters)}, 124(3):30004, dec 2018.

\bibitem{Adam2014Cooperative}
Adam Wysocki, Roland~G. Winkler, and Gerhard Gompper.
\newblock Cooperative motion of active brownian spheres in three-dimensional
  dense suspensions.
\newblock {\em {EPL} (Europhysics Letters)}, 105(4):48004, feb 2014.

\bibitem{Mognetti2013Living}
B.~M. Mognetti, A.~\ifmmode \check{S}\else \v{S}\fi{}ari\ifmmode~\acute{c}\else
  \'{c}\fi{}, S.~Angioletti-Uberti, A.~Cacciuto, C.~Valeriani, and D.~Frenkel.
\newblock Living clusters and crystals from low-density suspensions of active
  colloids.
\newblock {\em Phys. Rev. Lett.}, 111:245702, Dec 2013.

\bibitem{Stenhammar2014Phase}
Joakim Stenhammar, Davide Marenduzzo, Rosalind~J. Allen, and Michael~E. Cates.
\newblock Phase behaviour of active brownian particles: the role of
  dimensionality.
\newblock {\em Soft Matter}, 10:1489--1499, 2014.

\bibitem{Stenhammar2015Activity}
Joakim Stenhammar, Raphael Wittkowski, Davide Marenduzzo, and Michael~E. Cates.
\newblock Activity-induced phase separation and self-assembly in mixtures of
  active and passive particles.
\newblock {\em Phys. Rev. Lett.}, 114:018301, Jan 2015.

\bibitem{Kudrolli2008Swarming}
Arshad Kudrolli, Geoffroy Lumay, Dmitri Volfson, and Lev~S. Tsimring.
\newblock Swarming and swirling in self-propelled polar granular rods.
\newblock {\em Phys. Rev. Lett.}, 100:058001, Feb 2008.

\bibitem{Narayan2007Long}
Vijay Narayan, Sriram Ramaswamy, and Narayanan Menon.
\newblock Long-lived giant number fluctuations in a swarming granular nematic.
\newblock {\em Science}, 317(5834):105--108, 2007.

\bibitem{Julien2010Collective}
Julien Deseigne, Olivier Dauchot, and Hugues Chat\'e.
\newblock Collective motion of vibrated polar disks.
\newblock {\em Phys. Rev. Lett.}, 105:098001, Aug 2010.

\bibitem{Giomi2013Swarming}
L.~Giomi, N.~Hawley-Weld, and L.~Mahadevan.
\newblock Swarming, swirling and stasis in sequestered bristle-bots.
\newblock {\em Proceedings of the Royal Society A: Mathematical, Physical and
  Engineering Sciences}, 469(2151):20120637, 2013.

\bibitem{Liu2019Self}
Guannan Liu, Adam Patch, Fatmag\"ul Bahar, David Yllanes, Roy~D. Welch,
  M.~Cristina Marchetti, Shashi Thutupalli, and Joshua~W. Shaevitz.
\newblock Self-driven phase transitions drive myxococcus xanthus fruiting body
  formation.
\newblock {\em Phys. Rev. Lett.}, 122:248102, Jun 2019.

\bibitem{Liu11905}
Quan-Xing Liu, Arjen Doelman, Vivi Rottsch{\"a}fer, Monique de~Jager, Peter
  M.~J. Herman, Max Rietkerk, and Johan van~de Koppel.
\newblock Phase separation explains a new class of self-organized spatial
  patterns in ecological systems.
\newblock {\em Proceedings of the National Academy of Sciences},
  110(29):11905--11910, 2013.

\bibitem{Scholz2018Inertial}
Christian Scholz, Soudeh Jahanshahi, Anton Ldov, and Hartmut L{\"o}wen.
\newblock Inertial delay of self-propelled particles.
\newblock {\em Nature communications}, 9(1):5156, 2018.

\bibitem{RevModPhys.88.045006}
Clemens Bechinger, Roberto Di~Leonardo, Hartmut L\"owen, Charles Reichhardt,
  Giorgio Volpe, and Giovanni Volpe.
\newblock Active particles in complex and crowded environments.
\newblock {\em Rev. Mod. Phys.}, 88:045006, Nov 2016.

\bibitem{Hartmut2020InertialEffects}
Hartmut Löwen.
\newblock Inertial effects of self-propelled particles: From active brownian to
  active langevin motion.
\newblock {\em The Journal of Chemical Physics}, 152(4):040901, 2020.

\bibitem{mandal2019motility}
Suvendu Mandal, Benno Liebchen, and Hartmut L{\"o}wen.
\newblock Motility-induced temperature difference in coexisting phases.
\newblock {\em Physical Review Letters}, 123(22):228001, 2019.

\bibitem{Yuanjian2020Rototaxis}
Yuanjian Zheng and Hartmut L\"owen.
\newblock Rototaxis: Localization of active motion under rotation.
\newblock {\em Phys. Rev. Research}, 2:023079, Apr 2020.

\bibitem{Deblais2018Boundariess}
A.~Deblais, T.~Barois, T.~Guerin, P.~H. Delville, R.~Vaudaine, J.~S.
  Lintuvuori, J.~F. Boudet, J.~C. Baret, and H.~Kellay.
\newblock Boundaries control collective dynamics of inertial self-propelled
  robots.
\newblock {\em Phys. Rev. Lett.}, 120:188002, May 2018.

\bibitem{ferkinhoff1983key}
William~D Ferkinhoff and Ralph~W Gundersen.
\newblock {\em A key to the whirligig beetles of Minnesota and adjacent states
  and Canadian provinces ({C}oleoptera: {G}yrinidae)}.
\newblock Science Museum of Minnesota, 1983.

\bibitem{Bernd1980Behavioural}
Bernd Heinrich and F.~Daniel Vogt.
\newblock Aggregation and foraging behavior of whirligig beetles ({G}yrinidae).
\newblock {\em Behavioral Ecology and Sociobiology}, 7(3):179--186, 1980.

\bibitem{Vulinec1989Aggregation}
K.~Vulinec and M.~C. Miller.
\newblock Aggregation and predator avoidance in whirligig beetles
  ({C}oleoptera: {G}yrinidae).
\newblock {\em Journal of the New York Entomological Society}, 97(4):438--447,
  1989.

\bibitem{Romey2015A}
William~L. Romey and Alicia~R. Lamb.
\newblock Flash expansion threshold in whirligig swarms.
\newblock {\em PLOS ONE}, 10(8):1--12, 08 2015.

\bibitem{Romey2015B}
William~L. Romey, Amy~L. Smith, and Jerome Buhl.
\newblock Flash expansion and the repulsive herd.
\newblock {\em Animal Behaviour}, 110:171 -- 178, 2015.

\bibitem{voise2011capillary}
Jonathan Voise, Michael Schindler, J{\'e}r{\^o}me Casas, and Elie Rapha{\"e}l.
\newblock Capillary-based static self-assembly in higher organisms.
\newblock {\em Journal of The Royal Society Interface}, 8(62):1357--1366, 2011.

\bibitem{Philamore2015Row}
Hemma Philamore, Jonathan Rossiter, Andrew Stinchcombe, and Ioannis Ieropoulos.
\newblock Row-bot: An energetically autonomous artificial water boatman.
\newblock In {\em 2015 IEEE/RSJ International Conference on Intelligent Robots
  and Systems (IROS)}, pages 3888--3893. IEEE, 2015.

\bibitem{Jia2015Energy}
Xinghua Jia, Zongyao Chen, Andrew Riedel, Ting Si, William~R Hamel, and Mingjun
  Zhang.
\newblock Energy-efficient surface propulsion inspired by whirligig beetles.
\newblock {\em IEEE Transactions on Robotics}, 31(6):1432--1443, 2015.

\bibitem{Bokeon2017Design}
Bokeon Kwak and Joonbum Bae.
\newblock Design of hair-like appendages and comparative analysis on their
  coordination toward steady and efficient swimming.
\newblock {\em Bioinspiration {\&} Biomimetics}, 12(3):036014, may 2017.

\bibitem{Fish2003turning}
Frank~E. Fish and Anthony~J. Nicastro.
\newblock Aquatic turning performance by the whirligig beetle: constraints on
  maneuverability by a rigid biological system.
\newblock {\em Journal of Experimental Biology}, 206(10):1649--1656, 2003.

\bibitem{Xu2012Experimental}
Zhonghua Xu, Scott~C Lenaghan, Benjamin~E Reese, Xinghua Jia, and Mingjun
  Zhang.
\newblock Experimental studies and dynamics modeling analysis of the swimming
  and diving of whirligig beetles ({C}oleoptera: {G}yrinidae).
\newblock {\em PLoS computational biology}, 8(11):e1002792, 2012.

\bibitem{Voise2009Management}
Jonathan Voise and Jérôme Casas.
\newblock The management of fluid and wave resistances by whirligig beetles.
\newblock {\em Journal of the Royal Society, Interface / the Royal Society},
  7:343--52, 08 2009.

\bibitem{Damme2019Interparticle}
Robin van Damme, Jeroen Rodenburg, Ren{\'e} van Roij, and Marjolein Dijkstra.
\newblock Interparticle torques suppress motility-induced phase separation for
  rodlike particles.
\newblock {\em The Journal of chemical physics}, 150(16):164501, 2019.

\bibitem{Zhang2008Global}
{Li Zhang}, {Yuan Li}, and R.~{Nevatia}.
\newblock Global data association for multi-object tracking using network
  flows.
\newblock In {\em 2008 IEEE Conference on Computer Vision and Pattern
  Recognition}, pages 1--8, 2008.

\bibitem{trackingdraft}
S~Kumar, P~Nandakishore, H~L Devereux, C~R Twomey, A~Chakraborty, and
  S~Thutupalli.
\newblock Detecting and tracking high density homogeneous objects in low
  framerate videos.
\newblock {\em in preparation}, 2021.

\bibitem{schaap2007delaunay}
Willem~Egbert Schaap.
\newblock The delaunay tessellation field estimator.
\newblock {\em Ph. D. Thesis}, 2007.

\bibitem{Sosna20556}
Matthew M.~G. Sosna, Colin~R. Twomey, Joseph Bak-Coleman, Winnie Poel, Bryan~C.
  Daniels, Pawel Romanczuk, and Iain~D. Couzin.
\newblock Individual and collective encoding of risk in animal groups.
\newblock {\em Proceedings of the National Academy of Sciences},
  116(41):20556--20561, 2019.

\bibitem{bayes-opt}
Fernando Nogueira.
\newblock {Bayesian Optimization}: Open source constrained global optimization
  tool for {Python}, 2014--.

\bibitem{silverman1986density}
Bernard~W Silverman.
\newblock {\em Density estimation for statistics and data analysis}, volume~26.
\newblock CRC press, 1986.

\bibitem{Nickolls2008}
John Nickolls, Ian Buck, Michael Garland, and Kevin Skadron.
\newblock Scalable parallel programming with cuda.
\newblock {\em Queue}, 6(2):40–53, March 2008.

\end{thebibliography}
\end{document}